\documentclass[aps,pra,twocolumn,showpacs]{revtex4}
\usepackage{amsmath,amssymb,latexsym}
\usepackage[english]{babel}
\usepackage{graphicx}

\newtheorem{lemma}{Lemma}
\newtheorem{theorem}{Theorem}

\begin{document}

\title{Lower bounds on concurrence and separability conditions}
\author{Julio I.\ \surname{de Vicente}} 
\affiliation{Departamento de Matem\'aticas, Universidad Carlos III
de Madrid, Avda.\ de la Universidad 30, 28911 Legan\'es, Madrid,
Spain}

\begin{abstract}
We obtain analytical lower bounds on the concurrence of bipartite
quantum systems in arbitrary dimensions related to the violation of
separability conditions based on local uncertainty relations and on
the Bloch representation of density matrices. We also illustrate how
these results complement and improve those recently derived  [K.
Chen, S. Albeverio, and S.-M. Fei, Phys. Rev. Lett. \textbf{95},
040504 (2005)] by considering the Peres-Horodecki and the computable
cross norm or realignment criteria.
\end{abstract}

\pacs{03.67.Mn, 03.65.Ud}

\maketitle

\section{Introduction}

Entanglement 
is an essential ingredient in many applications of quantum
information theory such as dense coding, teleportation, quantum
cryptography and quantum computing \cite{Nie}. Therefore, the
characterization and quantification of entanglement are of great
importance in this field. However, and despite many efforts in the
last decade, a completely satisfactory solution to both problems has
not been found. Attending to the first one (the so-called
separability problem), there exist, nevertheless, several sufficient
conditions for the detection of entanglement. The most powerful is
known as the Peres-Horodecki or positive partial transpose (PPT)
criterion \cite{Per}, which is also necessary for low-dimensional
systems ($2\times2$ and $2\times3$) \cite{Hor1}. Another remarkable
sufficient condition is given by the computable cross norm
\cite{Rud} or realignment \cite{Che} (CCNR) criterion, which allows
to detect many entangled states for which the PPT criterion fails.
Recently, another criterion with this property \cite{deV} has been
developed by the author which we shall denote as correlation matrix
(CM) criterion. There are also other important criteria, which,
however, lack the operational character of the aforementioned ones,
since they are stated in terms of mean values or variances of
observables which have to be chosen wisely. This is the case of
conditions based on entanglement witnesses (EWs)
\cite{Hor1,EW} or uncertainty relations \cite{HT,ur}. 
In what comes to the quantification of entanglement, there exist a
large variety of proposed measures \cite{Ple}. However, the explicit
computation of these measures for arbitrary states is a very hard
task, not only analytically but also from the computational point of
view since they require optimization over a large number of
parameters \footnote{Except for the case of the negativity [G. Vidal
and R.F. Werner, Phys. Rev. A \textbf{65}, 032314 (2002)], which
relies on the Peres-Horodecki criterion and therefore fails to
quantify the entanglement of PPT entangled states.}. The only
measures for which an analytical expression is available is the
entanglement of formation \cite{EOF} and the concurrence
\cite{Woo,Run}, for which Wootters \cite{Woo} derived a formula in
the case of two qubits. Given the aforementioned difficulties for
the evaluation of the concurrence for higher dimensions, good bounds
for the estimation of this quantity have been sought. While, by
construction, upper bounds are numerically affordable, the
derivation of lower bounds has demanded a more thorough analysis
\cite{Min}. 
A completely analytical and powerful lower bound for the concurrence
was found in \cite{caf1} by relating this quantity with the PPT and
CCNR criteria, giving shape, therefore, to the intuitive idea that a
stronger violation of a separability condition may indicate a higher
amount of entanglement. In fact, a possible connection between the
value of concurrence and the violation of a separability condition
based on local uncertainty relations (LURs) was already suggested in
\cite{HT} (see \cite{SG,Kot} for further discussions on LURs and the
quantification of entanglement). The aim of this paper is to sharpen
the bounds of \cite{caf1} by relating concurrence and the LURs and
CM criteria, giving as a by-product a deeper insight in the above
idea. Partial improvements on these bounds have already been
achieved in the particular case of $N\times N$ quantum systems with
even $N\geq4$ in \cite{Bre1} by considering an EW based separability
criterion \cite{Bre2}, but the approach here is valid for the
general case. The bounds imposed by measurements of arbitrary EWs on
different entanglement measures have been recently studied in
\cite{new}.

\section{New lower bounds on concurrence}

We start by recalling the definitions of the concepts and quantities
we are dealing with. Let $H_A\simeq\mathbb{C}^M$ and
$H_B\simeq\mathbb{C}^N$ denote the Hilbert spaces of subsystems $A$
and $B$ ($M\leq N$). Then, the quantum state of the total system is
characterized by the density operator $\rho\in
\mathcal{B}(H_A\otimes H_B)$, where $\mathcal{B}(H)$ stands for the
real vector space of Hermitian operators acting on $H$, which is a
Hilbert-Schmidt space (with inner product
$\langle\rho,\tau\rangle_{HS}=\textrm{Tr}(\rho^\dag\tau)$). The
state is said to be separable (entangled) if it can (cannot) be
written as a convex combination of product states \cite{Wer}, i. e.
$\rho=\sum_i p_i \, \rho^A_i\otimes\rho^B_i$ where $0 \leq p_i \leq
1$, $\sum_i p_i = 1$, and $\rho_i^A$ ($\rho_i^B$) denotes a pure
state density matrix acting on $H_A$ ($H_B$). 
The generalized definition \cite{Run} for the concurrence of a pure
state $\psi$ is given by $C(\psi)=\sqrt{2(1-\textrm{Tr}\rho_A^2)}$,
where the reduced density matrix $\rho_A$ is obtained by tracing out
subsystem $B$ ($\rho_A=\textrm{Tr}_B|\psi\rangle\langle\psi|$).
Notice that $0\leq C(\psi)\leq \sqrt{2(M-1)/M}$, the lower bound
being attained by product states and the upper bound by maximally
entangled states. The definition is extended to general mixed states
$\rho$ by the convex roof, 
\begin{equation}\label{concurrence}
C(\rho)=\min_{\{p_i,|\psi_i\rangle\}}\left\{\sum_ip_iC(\psi_i) :
\rho=\sum_ip_i|\psi_i\rangle\langle\psi_i|\right\}.
\end{equation}
Consequently, $C(\rho)=0$ if, and only if, $\rho$ is a separable
state. The PPT and CCNR criteria can be formulated in several ways.
Basically, they state that certain rearrangements of the matrix
elements of $\rho$ \cite{Hor2}, namely, the partial transpose
$T_A(\rho)$ (PPT criterion) and the realignment $R(\rho)$ (CCNR
criterion), are such that for separable states $||T_A(\rho)||=1$ and
$||R(\rho)||\leq1$, where here, and throughout the paper,
$||\cdot||$ stands for the trace norm (i.e. the sum of the singular
values). By directly relating both criteria and concurrence by means
of the Schmidt coefficients of a pure state it was found in
\cite{caf1} that
\begin{equation}\label{caf}
C(\rho)\geq\sqrt{\frac{2}{M(M-1)}}\left[\max(||T_A(\rho)||,||R(\rho)||)-1\right].
\end{equation}

\subsection{LURs criterion}

One of the most interesting separability criteria based on
uncertainty relations is that of LURs \cite{HT}, since it can detect
PPT entanglement \cite{Hof,Guh}. It states that if $\{A_i\}$ and
$\{B_i\}$ are observables acting on $H_A$ and $H_B$ respectively,
fulfilling uncertainty relations $\sum_i\Delta^2_{\rho}(A_i)\geq
C_A$ and $\sum_i\Delta^2_{\rho}(B_i)\geq C_B$ ($C_A,C_B\geq0$),
then,
\begin{equation}\label{LUR}
\sum_i\Delta^2_\rho(A_i\otimes I+I\otimes B_i)\geq C_A+C_B
\end{equation}
holds for separable states \cite{HT}. The variance $\Delta^2$ is
given by $\Delta_\rho^2(M)=\langle M^2\rangle_\rho-\langle
M\rangle^2_\rho$, where $\langle M\rangle_\rho=\textrm{Tr}(\rho M)$
is the expectation value of the observable $M$. A particularly
interesting choice for the observables is that of local orthogonal
observables (LOOs) \cite{Loo}, that is, orthonormal bases of
$\mathcal{B}(H_A)$ and $\mathcal{B}(H_B)$, which we shall denote
$\{G_i^A\}_{i=1}^{M^2}$ and $\{G_i^B\}_{i=1}^{N^2}$. In this case
Eq.\ (\ref{LUR}) reads \cite{Guh}
\begin{equation}\label{LURLOOs}
\sum_{i=1}^{N^2}\Delta^2_\rho(G^A_i\otimes I+I\otimes G^B_i)\geq
M+N-2
\end{equation}
since
\begin{equation}\label{urLOOs}
\sum_{i=1}^{M^2}\Delta^2_\rho(G^A_i)\geq
M-1,\quad\sum_{i=1}^{N^2}\Delta^2_\rho(G^B_i)\geq N-1.
\end{equation}
Notice that if $M<N$, in Eq.\ (\ref{LURLOOs}) (and throughout the
paper) it is understood that $G_i^A=0$ for $M^2+1\leq i\leq N^2$.
The standard set of LOOs is given by
$\{G_i^A\}=\{g_j,g_{jk}^+,g_{jk}^-\}$ where
\begin{align}
g_j &= |j\rangle\langle j| \quad (0\leq j\leq M-1),\nonumber\\
g_{jk}^+ &=\frac{1}{\sqrt{2}}(|j\rangle\langle k|+|k\rangle\langle
j|)\quad (0\leq j<k\leq M-1),\\
g_{jk}^- &=-\frac{i}{\sqrt{2}}(|j\rangle\langle k|-|k\rangle\langle
j|)\quad (0\leq j<k\leq M-1),\nonumber
\end{align}
and similarly for $\{G_i^B\}$. The importance of the LURs condition
formulated in terms of LOOs relies on that it is strictly stronger
than the CCNR condition \cite{Guh}. Furthermore, it can detect
entangled states for which both the PPT and CCNR criteria fail
\cite{Guh}. In order to relate concurrence and LURs with LOOs
analogously as in Eq.\ (\ref{caf}) we start with the following
lemma:
\begin{lemma}
For any set of LOOs $\{G_i^A\}$ and $\{G_i^B\}$ and any $M\times N$
($M\leq N$) pure state $\psi$ with Schmidt decomposition
$|\psi\rangle=\sum_{j=0}^{M-1}\sqrt{\mu_j}|j_Aj_B\rangle$,
\begin{equation}\label{lema}
\sum_{i=1}^{N^2}\Delta^2_\psi(G^A_i\otimes I+I\otimes G^B_i)\geq
M+N-2-4\sum_{j<k}\sqrt{\mu_j\mu_k}
\end{equation}
holds. The bound is attained when
$\{G_i^A\}=\{g_j,g_{jk}^+,g_{jk}^-\}$ and
$\{G_i^B\}=\{-g_j,-g_{jk}^+,g_{jk}^-\}$ (constructed from the
corresponding Schmidt basis).
\end{lemma}
\textit{Proof.} We have that
\begin{align*}
\sum_i\Delta^2_\psi(G^A_i\otimes I+I\otimes G^B_i)&=
\sum_i\left(\Delta^2_{\rho_A}(G_i^A)
+\Delta^2_{\rho_B}(G_i^B)\right)\\&+2\sum_i\kappa_\psi(G^A_i,G^B_i),
\end{align*}
where
\begin{equation*}
\kappa_\psi(G^A_i,G^B_i)=\langle G_i^A\otimes
G_i^B\rangle_\psi-2\langle G_i^A\rangle_{\rho_A}\langle
G_i^B\rangle_{\rho_B}.
\end{equation*}
Let us write
$\rho_\psi=|\psi\rangle\langle\psi|=\rho^{sep}+\epsilon$, where
$\rho^{sep}=\sum_{j}\mu_j|j_Aj_B\rangle\langle j_Aj_B|$ and
$\epsilon=\sum_{j\neq k}\sqrt{\mu_j\mu_k}|j_Aj_B\rangle\langle
k_Ak_B|$. Notice that $\rho^{sep}$ is a separable state and that its
reductions are the same as those of $\rho_\psi$
($\rho^{sep}_{A}=\rho_{A}$, $\rho^{sep}_{B}=\rho_{B}$). Thus,
\begin{align*}
\kappa_\psi(G^A_i,G^B_i) & =\textrm{Tr}(\epsilon G_i^A\otimes
G_i^B)+\kappa_{\rho^{sep}}(G^A_i,G^B_i).
\end{align*}
Now, since LURs hold for separable states we have that
\begin{align*}
2\sum_i\kappa_{\rho^{sep}}(G^A_i,G^B_i)&\geq
M+N-2\\&-\sum_i\left(\Delta^2_{\rho_A}(G_i^A)
+\Delta^2_{\rho_B}(G_i^B)\right)
\end{align*}
and then,
\begin{equation*}
\sum_i\Delta^2_\psi(G^A_i\otimes I_B+I_A\otimes G^B_i)\geq
M+N-2+2\sum_i\textrm{Tr}(\epsilon G_i^A\otimes G_i^B),
\end{equation*}
so that it remains to prove that $X\equiv\sum_i\textrm{Tr}(\epsilon
G_i^A\otimes G_i^B)\geq-2\sum_{j<k}\sqrt{\mu_j\mu_k}$. To do so,
notice that
\begin{align}
X&=\sum_i\sum_{j\neq k}\sqrt{\mu_j\mu_k}\langle
j_A|G_i^A|k_A\rangle\langle j_B|G_i^B|k_B\rangle\nonumber\\
& \geq-\sum_i \sum_{j\ne k} \sqrt{\mu_j\mu_k}|\langle
j_A|G_i^A|k_A\rangle\langle j_B|G_i^B|k_B\rangle| \nonumber \\ &\ge
-\sum_i \sum_{j\ne k} \sqrt{\mu_j\mu_k}(|\langle
j_A|G_i^A|k_A\rangle|^2+|\langle
j_B|G_i^B|k_B\rangle|^2)/2,\nonumber
\end{align}
where in the last step we have used that $a^2+b^2\geq2|ab|$. Now,
the result follows because $\sum_i\left|\langle
j_A|G_i^A|k_A\rangle\right|^2=\sum_i\left|\langle
j_B|G_i^B|k_B\rangle\right|^2=1$ $\forall j,k$ for any set of LOOs
$\{G_i^A\}$ and $\{G_i^B\}$. To see this, consider that
$\mathcal{B}(H_A)$ is isomorphic to $\mathbb{C}^{M^2}$ with the
standard inner product, so that the $\{G_i^A\}$ can be arranged as
column vectors which give an orthonormal basis of this space. This
column vectors together give rise to a unitary matrix $U$, and
$\sum_i\left|\langle j_A|G_i^A|k_A\rangle\right|^2$ corresponds to
summing the squared modulus of the elements of a certain row of $U$
and, therefore, it equals unity. Obviously the same reasoning holds
for $\sum_i\left|\langle j_B|G_i^B|k_B\rangle\right|^2$. It remains
to check that the bound is attained by the above stated set of LOOs.
Using that $\sum_i(G_i^A)^2=MI$ and $\sum_i(G_i^B)^2=NI$ \cite{Guh},
it is straightforward to find that
\begin{align}
\sum_{i=1}^{N^2}\Delta^2_\psi(G^A_i\otimes I+I\otimes
G^B_i)&=M+N+2\sum_i\langle G_i^A\otimes
G_i^B\rangle_\psi\nonumber\\
& -\sum_i\langle G^A_i\otimes I+I\otimes
G^B_i\rangle^2_\psi.\nonumber
\end{align}
Considering that any pure density matrix $\rho_\psi$ achieves its
Schmidt decomposition for the standard LOOs, i.e.
$\rho_\psi=\sum_j\mu_jg_j\otimes
g_j+\sum_{j<k}\sqrt{\mu_j\mu_k}(g_{jk}^+\otimes
g_{jk}^++g_{jk}^-\otimes(-g_{jk}^-))$, it follows that
$\sum_i\langle G_i^A\otimes
G_i^B\rangle_\psi=-1-2\sum_{j<k}\sqrt{\mu_j\mu_k}$ and that $\langle
G^A_i\otimes I+I\otimes G^B_i\rangle_\psi=0$ $\forall i$ for the set
of LOOs mentioned in the statement of the lemma, and the result is
thus proved.

Now, we can prove our main result.
\begin{theorem}
For any $M\times N$ ($M\leq N$) quantum state $\rho$,
\begin{equation}\label{t1}
C(\rho)\geq\frac{M+N-2-\sum_i\Delta^2_\rho(G^A_i\otimes I+I\otimes
G^B_i)}{\sqrt{2M(M-1)}}
\end{equation}
holds for any set of LOOs $\{G_i^A\}$ and $\{G_i^B\}$.
\end{theorem}
\textit{Proof.} Let $\sum_np_n|\psi_n\rangle\langle\psi_n|$ be the
decomposition of $\rho$ for which the minimum in Eq.\
(\ref{concurrence}) is attained, so that,
$C(\rho)=\sum_np_nC(\psi_n)$. Since the concurrence of a pure state
is directly related to its Schmidt coefficients \cite{Run}:
$C^2(\psi_n)=4\sum_{j<k}\mu_j\mu_k$, and \cite{caf1}
\begin{equation}\label{desigualdad}
\sum_{j<k}\mu_j\mu_k\geq\frac{2}{M(M-1)}(\sum_{j<k}\sqrt{\mu_j\mu_k})^2,
\end{equation}
we have that
$C(\psi_n)\geq\sqrt{2/(M(M-1))}2\sum_{j<k}\sqrt{\mu_j\mu_k}$. Now,
the use of Lemma 1 and the fact that
$\Delta^2_\rho(M)\geq\sum_np_n\Delta^2_{\psi_n}(M)$ for any
observable $M$ (see e.g. \cite{HT}) proves the desired result.

Next we present a couple of examples which illustrate how this
result can improve on the bounds of \cite{caf1}. First, consider the
case of PPT entangled states. In this case any non-trivial bound on
concurrence given by Eq.\ (\ref{caf}) must rely on the CCNR
criterion. However, the LURs criterion can identify states of this
kind which are not detected by the CCNR criterion and, therefore,
place a non-trivial bound on concurrence where the previous approach
failed (see \cite{Guh}). Furthermore, Eq.\ (\ref{t1}) can improve
the estimation of  PPT entanglement of Eq.\ (\ref{caf}) even when
the latter supplies a non-trivial bound. For instance, consider the
following $3\times3$ PPT entangled state constructed in \cite{upb1}
from unextendible product bases (UPB):
$\rho=1/4(I-\sum_i|\psi_i\rangle\langle\psi_i|)$, where
$|\psi_0\rangle=|0\rangle(|0\rangle-|1\rangle)/\sqrt{2}$,
$|\psi_1\rangle=(|0\rangle-|1\rangle)|2\rangle/\sqrt{2}$,
$|\psi_2\rangle=|2\rangle(|1\rangle-|2\rangle)/\sqrt{2}$,
$|\psi_3\rangle=(|1\rangle-|2\rangle)|0\rangle/\sqrt{2}$ and
$|\psi_4\rangle=(|0\rangle+|1\rangle+|2\rangle)(|0\rangle+|1\rangle+|2\rangle)/3$.
While Eq.\ (\ref{caf}) yields $C(\rho)\geq0.050$ \cite{caf1},
Theorem 1 with the LOOs used in \cite{Guh} to improve the detection
of $\rho$ mixed with white noise gives $C(\rho)\geq0.052$. Another
interesting example is to consider the $2\times3$ state
$\varrho=p|\Psi\rangle\langle\Psi|+(1-p)|01\rangle\langle01|$, where
$|\Psi\rangle=(|00\rangle+|11\rangle)/\sqrt{2}$, which shows that
(\ref{t1}) can give a better bound for the concurrence than
(\ref{caf}) even though the PPT criterion characterizes entanglement
optimally in this case (see Fig. 1).

\begin{figure}[t]
\includegraphics[scale=0.4]{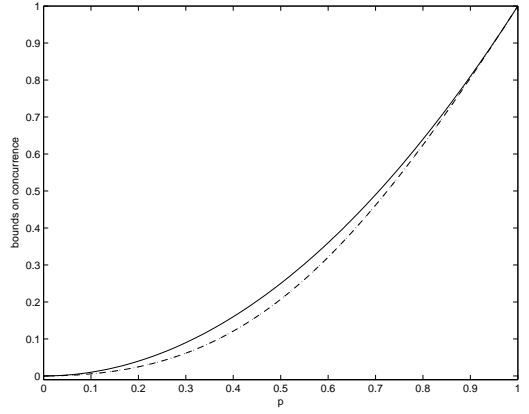}
\caption{Lower bounds on concurrence for the the state $\varrho$.
Solid line: Bound given by the LURs criterion with the LOOs that
achieve the bound in Eq. (\ref{lema}) for the state $\Psi$. Dotted
line: Bound given by the PPT criterion. Dashed line: Bound given by
the CCNR criterion. (The last two bounds overlap).}
\end{figure}

\subsection{CM criterion}

Besides its PPT entanglement detection capability, the CM criterion
can be stronger than the CCNR criterion when $M\neq N$ \cite{deV},
so it seems worthwhile to derive an analogous result relying on this
criterion. It is based on the Bloch representation of density
matrices which is achieved by expanding $\rho$ with respect to a
particular set of unnormalized LOOs, namely, the identity and the
traceless Hermitian generators of $SU(M)$ and $SU(N)$ (denoted
$\{\lambda_i^A\}_{i=1}^{M^2-1}$ and $\{\lambda_i^B\}_{i=1}^{N^2-1}$
hereafter), i.e. $\rho=1/MN(I\otimes I+\sum_ir_i\lambda^A_i\otimes
I+\sum_js_jI\otimes\lambda^B_j+\sum_{i,j}t_{ij}\lambda_i^A\otimes\lambda^B_j)$.
The generators $\{\lambda_i^A\}=\{w_l,u_{jk},v_{jk}\}$ of $SU(M)$
can be constructed from any orthonormal basis in $H_A$ \cite{Hio},
\begin{align}
& w_l=\sqrt{\frac{2}{(l+1)(l+2)}}\left(\sum_{i=0}^l|i\rangle\langle
i|-(l+1)|l+1\rangle\langle l+1|\right),\nonumber\\
& u_{jk}=\sqrt{2}g_{jk}^+,\quad v_{jk}=\sqrt{2}g_{jk}^-,
\end{align}
where $0\leq l\leq M-2$ and $0\leq j<k\leq M-1$. The Bloch
representation has two kind of parameters: $\{r_i\}$ and $\{s_i\}$,
which are local since they are the Bloch parameters of the
reductions ($\rho_A=1/M(I+\sum_ir_i\lambda_i^A)$,
$\rho_B=1/N(I+\sum_is_i\lambda^B_i)$); and
$\{t_{ij}\}=MN/4\{\langle\lambda_i^A\otimes\lambda_j^B\rangle_\rho\}$,
which are responsible for the possible correlations between the
subsystems. These last coefficients can be arranged to form the CM,
$(T)_{ij}=t_{ij}$. The CM criterion affirms that there is an upper
bound to the correlations inherent in a separable state since
$||T||\leq K_{MN}=\sqrt{MN(M-1)(N-1)}/2$ must hold for these states
\cite{deV}.
\begin{theorem}
For any $M\times N$ ($M\leq N$) quantum state $\rho$,
\begin{equation}\label{t2}
C(\rho)\geq\sqrt{\frac{8}{M^3N^2(M-1)}}\left(||T||-K_{MN}\right)
\end{equation}
holds.
\end{theorem}
\textit{Proof.} As before, let us first relate $||T_\psi||$ of a
pure state to its Schmidt coefficients. Following the notation of
the proof of Lemma 1 we write the pure state density matrix as
$\rho_\psi=\rho^{sep}+\epsilon=\rho^{sep}+1/2\sum_{j<k}\sqrt{\mu_j\mu_k}(u_{jk}\otimes
u_{jk}-v_{jk}\otimes v_{jk})$. Since $\rho^{sep}$ is diagonal, its
Bloch representation is just given in terms of the identity and the
$w_l$'s. Therefore, the CM of $\rho_\psi$ is block-diagonal and,
thus,
$||T_\psi||=||T_{\rho^{sep}}||+MN\sum_{j<k}\sqrt{\mu_j\mu_k}\leq
K_{MN}+MN\sum_{j<k}\sqrt{\mu_j\mu_k}$. Hence, using again Eq.\
(\ref{desigualdad}) we have that
$$C(\psi)\geq\sqrt{\frac{8}{M^3N^2(M-1)}}\left(||T_\psi||-K_{MN}\right).$$
Let $\sum_np_n|\psi_n\rangle\langle\psi_n|$ denote the ensemble
decomposition of $\rho$ for which $C(\rho)=\sum_np_nC(\psi_n)$.
Then, we can use the above inequality for every $\psi_n$ together
with the triangle inequality
($||T_\rho||=||\sum_np_nT_{\psi_n}||\leq\sum_np_n||T_{\psi_n}||$) to
prove the claim.

Regrettably, to find examples in which Eq.\ (\ref{t2}) improves the
bound given by Eq.\ (\ref{caf}) is harder than in the case of LURs.
This is, among other reasons, because the norm of the CM of a pure
state is related to the Schmidt coefficients through an inequality,
while in the PPT and CCNR cases this kind of relation was given by
equality. Thus, Theorem 2 is only expected to improve on the result
of \cite{caf1} for states which are detected by the CM criterion but
not by the PPT and CCNR criteria, or, more generally, in situations
where the former criterion is stronger than both the later criteria
at the same time \cite{inpreparation}.

\section{Conclusions}

We have derived an analytical lower bound for the concurrence
related to the LURs criterion for separability. We have shown by
considering explicit examples how this result can improve the bounds
given in \cite{caf1}, which rely on the PPT and CCNR criteria. We
have also shown that this new result can yield better bounds for the
estimation of concurrence even in situations where the PPT criterion
is optimal for the detection of entanglement. However, Eq.\
(\ref{t1}) should not be considered to render Eq.\ (\ref{caf})
obsolete but rather as a complement of it that can be used to refine
the bounds that (\ref{caf}) provides when a suitable choice of LOOs
is made. To determine what set of LOOs yields the best bound for a
given state is left as an interesting open problem. We also think
that this result helps to understand the relation between
entanglement quantification and the LURs criterion. Like the results
of \cite{caf1} our bound can be attained by states having a
particular optimal ensemble decomposition.
This is the case of isotropic states \cite{iso} when
$\{G_i^A\}=\{I/\sqrt{M},w_l/\sqrt{2},u_{jk}/\sqrt{2},v_{jk}/\sqrt{2}\}$
and
$\{G_i^B\}=\{-I/\sqrt{N},-w_l/\sqrt{2},-u_{jk}/\sqrt{2},v_{jk}/\sqrt{2}\}$
are the chosen LOOs, and, hence, this explains the coincidence of
concurrence and violation of LURs pointed out in \cite{HT}.

We have also provided a similar lower bound on concurrence in terms
of violations of the CM criterion for separability. Although this
result is not as powerful as the one based on LURs, since it only
seems to yield better bounds than what can be obtained using Eq.\
(\ref{caf}) in situations where the CM criterion has a stronger
entanglement detection capability than both the PPT and CCNR
criteria jointly, it provides a rigorous relation between
concurrence and a correlation-based local unitary invariant measure,
which is convenient from the experimental point of view as discussed
in \cite{Kot}.

Finally, let us mention that the results presented here can be
extended straightforwardly to yield lower bounds for the
entanglement of formation by using the ideas of \cite{caf2}.

\section*{ACKNOWLEDGMENTS}

The author acknowledges financial support by Universidad Carlos III
de Madrid and Comunidad Aut\'onoma de Madrid (project No.
CCG06-UC3M/ESP-0690) and by Direcci\'on General de Investigaci\'on
(Ministerio de Educaci\'on y Ciencia of Spain) under grant
MTM2006-13000-C03-02.


\end{document}